\documentclass[12pt]{iopart}
\usepackage{epsfig,amssymb,psfig}

\def\spose#1{\hbox to 0pt{#1\hss}}
\def\lesssim{\mathrel{\spose{\lower 3pt\hbox{$\mathchar"218$}}
 \raise 2.0pt\hbox{$\mathchar"13C$}}}
\def\gtrsim{\mathrel{\spose{\lower 3pt\hbox{$\mathchar"218$}}
 \raise 2.0pt\hbox{$\mathchar"13E$}}}

\def\<{\langle}
\def\>{\rangle}

\begin{document}

\title{ Quasi-long-range order in 
the 2D XY model with random phase shifts  }

\author{Vincenzo Alba$^1$, Andrea Pelissetto$^2$ and Ettore Vicari$^3$}
\address{$^1$ Scuola Normale Superiore and INFN, I-56126 Pisa, Italy}
\address{$^2$ Dipartimento di Fisica dell'Universit\`a di Roma ``La Sapienza"
        and INFN, I-00185 Roma, Italy}
\address{$^3$ Dipartimento di Fisica dell'Universit\`a di Pisa
        and INFN, I-56127 Pisa, Italy}

\ead{
Vincenzo.Alba@sns.it,
Andrea.Pelissetto@roma1.infn.it,
Ettore.Vicari@df.unipi.it}

\begin{abstract}
  We study the square-lattice XY model in the presence of random phase 
  shifts. We consider two different disorder distributions
  with zero average shift and 
  investigate the low-temperature quasi-long-range order phase 
  which occurs for sufficiently low disorder. By means of 
  Monte Carlo simulations we determine several universal quantities
  which are then compared with the analytic predictions of the 
  random spin-wave theory. We observe a very good agreement which indicates 
  that the universal long-distance behavior in the whole low-disorder 
  low-temperature phase is fully described by the random spin-wave theory.

\end{abstract}

%%%%% \pacs{PACS Numbers: 74.81.Fa, 05.70.Jk, 64.60.Fr}
%74.81.Fa Josephson junction arrays
%64.60.-i general studies of phase transitions
%64.60.Fr Equilibrium properties near critical points
%75.10.Hk classical spin models
%74.78.-w       Superconducting films and low-dimensional structures
%05.10.Ln       Monte Carlo methods
%05.70.Jk       Critical point phenomena 

\maketitle

% ========================= BODY =========================
%\narrowtext

\section{Introduction}
\label{intro}

The two-dimensional XY model with random phase shifts (RPXY) describes the
thermodynamic behavior of several disordered systems, such as magnetic systems
with random Dzyaloshinskii-Moriya interactions~\cite{RSN-83}, Josephson
junction arrays with geometrical disorder~\cite{GK-86,GK-89}, crystal systems
on disordered substrates~\cite{CF-94}, and vortex glasses~\cite{FTY-91}.  See
\cite{Korshunov-06,KR-03} for recent reviews.  The RPXY model is defined
by the Hamiltonian
\begin{eqnarray}
{\cal H} = -\sum_{\langle xy \rangle } {\rm Re} \,\psi_x^* U_{ij} \psi_y
= - \sum_{\langle xy \rangle} {\rm cos}(\theta_x - \theta_y-A_{xy}),
\label{RPXY} 
\end{eqnarray}
where $\psi_x\equiv e^{i\theta_x}$, $U_{xy}\equiv e^{i A_{xy}}$, and the
sum runs over the bonds ${\langle xy \rangle }$ of a square lattice.  The
phases $A_{xy}$ are uncorrelated quenched random variables with zero
average. A Gaussian distribution,
\begin{equation}
P(A_{xy}) \propto \exp\left(-{A_{xy}^2\over 2\sigma}\right),
\label{paxy}
\end{equation}
has been considered in most of the studies of the RPXY.  We denote the
RPXY with Gaussian-distributed phases by GRPXY.  The pure XY model is recovered in the limit
$\sigma\rightarrow 0$, while the so-called gauge glass model~\cite{ES-85} with
uniformly distributed phase shifts is obtained in the limit $\sigma\rightarrow
\infty$.

The pure XY model shows a high-temperature paramagnetic phase and a
low-temperature phase characterized by  quasi-long-range order (QLRO) controlled by a
line of Gaussian fixed points. In the latter phase, the spin-spin correlation 
function $\langle \bar{\psi}_x \psi_y \rangle$ decays as $1/|x-y|^{\eta(T)}$ 
for large $|x-y|$, with $\eta$ depending on $T$; $\eta\sim T$ for small values of $T$.  
The two phases
are separated by a Kosterlitz-Thouless (KT) transition~\cite{KT-73} at
\cite{HP-97} $\beta_{XY}\equiv 1/T_{XY}=1.1199(1)$. For 
$\tau \equiv T/T_{XY}-1\rightarrow 0^+$
the correlation
length diverges exponentially as ${\rm ln} \xi \sim \tau^{-1/2}$ 
and the magnetic susceptibility behaves as
$\chi\sim \xi^{7/4}$, corresponding to $\eta=1/4$.  

\begin{figure*}[tb]
\centerline{\psfig{width=9truecm,angle=0,file=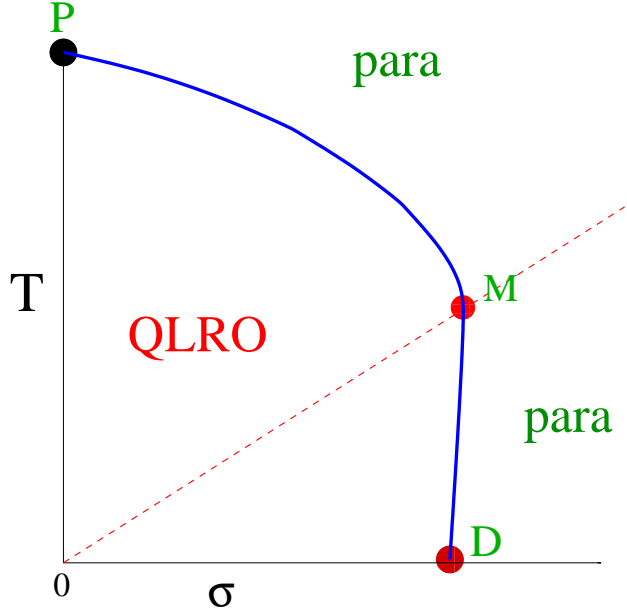}}
\caption{
Generic phase diagram of two-dimensional RPXY models 
in the temperature-disorder plane. 
The full line corresponds to the transition line. The dashed line
is the N line $T=\sigma$ and refers only to the CRPXY model.
}
\label{phadia}
\end{figure*}

In this paper we shall discuss the low-temperature 
behavior of RPXY models for small disorder.
It has already been investigated in several works, 
most of them focussing on the GRPXY; see,
e.g.,  \cite{RSN-83,ES-85,GK-86,FLTL-88,CD-88,GK-89,%%
  FBL-90,HS-90,RTYF-91,FTY-91,Li-92,Gingras-92,DWKHG-92,ON-93,RY-93,%%
  Korshunov-93,NK-93,Nishimori-94,CF-94,NSKL-95,CF-95,%%
  JKC-95,HWFGY-95,KN-96,Tang-96,MG-97,%%
  Scheidl-97,KS-97,KCRS-97,MG-98,CL-98,KS-98,MW-99,CP-99,%%
  Kim-00,CL-00,AK-02,KY-02,HKM-03,Katzgraber-03,KC-05}.
The expected $T$-$\sigma$ phase diagram, which is
sketched in Fig.~\ref{phadia}, presents two phases at finite temperature: a
paramagnetic and a QLRO phase.  The paramagnetic phase is separated from
the QLRO phase by a transition line, which starts from the pure XY point
(denoted by $P$ in Fig.~\ref{phadia}) at $(\sigma=0,T=T_{XY}\approx 0.893)$
and ends at a zero-temperature disorder-induced transition denoted by $D$ at
$(\sigma_0,T=0)$. Note that QLRO is observed only up to a maximum
value $\sigma_M$ of the disorder parameter, which is related to the 
point $M\equiv (\sigma_M,T_M)$, where 
the tangent to the transition line is parallel to the $T$ axis.
The transition line from $M$ to $D$ is believed to run (almost) parallel to the
$T$ axis, with $\sigma_0\lesssim \sigma_M$; see, e.g.,
\cite{Korshunov-06}.~\footnote{The first renormalization-group (RG) analyses
  based on a Coulomb-gas representation of the models~\cite{RSN-83} predicted reentrant
  transitions for any value of $\sigma\lesssim \pi/8$, so that $\sigma_0 = 0$. 
  It was then clarified that this was an artefact of the approximations.
  Indeed, in agrement with the experimental findings \cite{FLTL-88}, numerical
  works~\cite{FLTL-88,CD-88,FBL-90,MG-97} and more careful RG analyses
  \cite{ON-93,NSKL-95,KN-96,Tang-96,Scheidl-97,CL-98} showed the
  absence of reentrant transitions for sufficiently small values of
  $\sigma$, that is predicted $\sigma_0 > 0$. For the CRPXY, \cite{ON-93} suggested
  that the transition line is parallel to the $T$ axis below the Nishimori
  point $M$. 
  It is not clear whether this conjecture is correct. Indeed,
  the analogous conjecture fails in the case of the 2D $\pm J$ Ising model;
  see \cite{PPV-08} and references therein.}

The QLRO phase of the pure XY model is expected to survive for sufficiently
small values of $\sigma$ (see, however, \cite{MW-99} for a recent critical
discussion of this scenario).  It disappears for large disorder, 
for~\cite{RSN-83} $\sigma\gtrsim \sigma_M\approx \pi/8$.  The RPXY model 
for large disorder, and in
particular the gauge-glass model ($\sigma\rightarrow \infty$), has
been much investigated~\cite{FTY-91,ES-85,HS-90,RTYF-91,Li-92,Gingras-92,%%
  DWKHG-92,RY-93,NK-93,Nishimori-94,JKC-95,HWFGY-95,KS-97,%%
  KCRS-97,MG-98,KS-98,Granato-98,CP-99,Kim-00,AK-02,KY-02,%%
  HKM-03,Katzgraber-03,KC-05}.  No long-range glassy order can exist at finite
temperature~\cite{NK-93,Nishimori-94}.  Some numerical works support a 
zero-temperature transition; see, e.g., \cite{Katzgraber-03,KY-02,AK-02}.

Beside the GRPXY, we also consider a RPXY model with a slightly different
distribution, given by
\begin{equation}
P(A_{xy})\propto \exp\left({{\rm cos} A_{xy}\over \sigma}\right).
\label{paxy2}
\end{equation}
We denote the RPXY model with the distribution (\ref{paxy2}) by CRPXY.  The
CRPXY is interesting because it allows to obtain exact results along the
so-called Nishimori (N) line~\cite{ON-93}
\begin{equation}
T =\sigma,
\label{Nline}
\end{equation}
exploiting gauge invariance~\cite{Nishimori-81,ON-93,Nishimori-02}.  For
example, the energy density is exactly known along the N line:
$E=-I_1(\beta)/I_0(\beta)$, where $\beta\equiv 1/T$ and $I_n(x)$ are the modified Bessel
functions.
Another important feature of the CRPXY is that along
the $N$ line the spin-spin and overlap correlation functions are equal, 
\begin{eqnarray}
[\langle {\psi}_x^* \psi_y \rangle ] = 
[|\langle {\psi}_x^* \psi_y \rangle|^2 ],
\label{cnl}
\end{eqnarray}
where the angular and square brackets indicate the
thermal average and the quenched average over disorder, respectively.  The phase diagram is
expected to be analogous to that of GRPXY model.  As already noted in
\cite{Nishimori-02}, the N line must play an important role in the phase
diagram, because it is expected to mark the crossover between the region
dominated by magnetic correlations and that dominated by randomness.  In
\cite{ON-93} it was proven that the critical value $\sigma_M$ of $\sigma$
along the N line is an upper bound for the values of $\sigma$ where 
magnetic QLRO can exist (note that this does not exclude the existence of 
a glassy QLRO for
$\sigma>\sigma_M$).  Therefore, we can identify it as the point $M$ where
$d\sigma_c/d T=0$ in Fig.~\ref{phadia}.  The critical value $\sigma_0$ at
$T=0$ must satisfy $\sigma_0\le\sigma_M$, leaving open the possibility of
reentrant transitions.

In this paper we focus on the QLRO phase of RPXY models, and investigate its
nature.  In particular, we check the random-spin-wave scenario~\cite{RSN-83},
in which the long-distance behavior is essentially identical to that in the 
model obtained by replacing
\begin{equation}
{\rm cos}(\theta_x - \theta_y-A_{xy})\longrightarrow
1 - {1\over 2} (\theta_x - \theta_y-A_{xy})^2
\label{swr}
\end{equation}
in the Hamiltonian (\ref{RPXY}).  We numerically check this scenario by
perfoming Monte Carlo (MC) simulations of the GRPXY and CRPXY models and 
provide conclusive
evidence of the existence of a low-$T$ phase with QLRO determined by the random
spin-wave theory.  Some numerical evidence of QLRO in the
GRPXY model was already presented in \cite{MG-97}.

The paper is organized as follows. In Sec.~\ref{rswt} we summarize the 
main predictions of the spin-wave theory, which are then compared with 
the numerical data in Sec.~\ref{mcres}. Our conclusions are presented in
Sec.~\ref{conclusions}. The definitions of the quantities we consider
are reported in \ref{notations}. In \ref{App:etasCPRXY} we report the spin-wave
calculation of $\eta_s$ in the CRPXY model.

\section{The random spin-wave theory}
\label{rswt}

In the spin-wave limit the partition function is given by
\begin{eqnarray}
Z(\{A\}) = \int [d\phi]  e^{- H_{sw}/T},\qquad
H_{sw} = {1\over 2} \sum_{\langle ij \rangle} (\phi_i-\phi_{j}-A_{ij})^2,
\label{swmod}
\end{eqnarray}
where the link-variables $A_{ij}$ are uncorrelated quenched random variables with
Gaussian probability $P(A_{ij})$.  For the GRPXY 
the spin and the overlap correlation functions are given
by~\cite{RSN-83,Nishimori-02}
\begin{eqnarray}
&&G_s(x-y)=[\langle e^{i(\phi_x-\phi_y)} \rangle] 
= \exp[(T+\sigma)G(x-y)], 
\label{sw2g} \\
&&G_o(x-y)=[|\langle e^{i(\phi_x-\phi_y)} \rangle|^2] = \exp [2TG(x-y)] ,
\label{sw2go} 
\end{eqnarray}
where $G(x-y)$ is the (infrared-regularized) two-point function of the Gaussian model without
disorder
\begin{equation}
G(r) = \int {d^2p\over (2\pi)^2} {e^{ip\cdot r} - 1\over p^2}~.
\label{Gaussian-prop}
\end{equation}
For the CRPXY one should take into account the nontrivial dependence of 
$P(A_{ij})$ on $A_{ij}$. For the overlap correlation function one still
obtains (\ref{sw2go}): for any probability distribution $G_o(x-y)$ does not
depend on randomness in the spin-wave approximation. 
For the spin correlation function the $\sigma$ dependence at $T=0$ is more 
complex. For $\sigma\to 0$ we obtain 
\begin{eqnarray}
G_s(x-y)
= \exp[(T+\sigma+\sigma^2/2)G(x-y)], 
\label{GsCPRXY}
\end{eqnarray}
disregarding terms of order $\sigma^3$.  
The derivation is reported in \ref{App:etasCPRXY}.

\begin{figure}[tb]
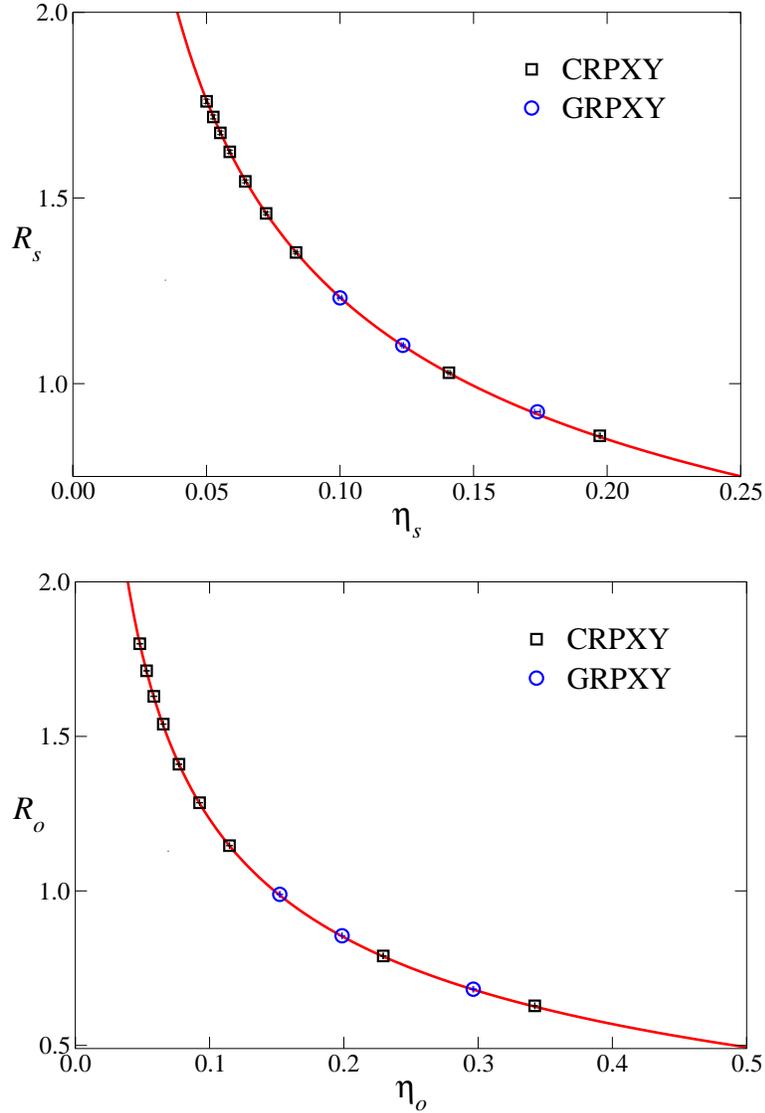

\centerline{\psfig{width=10truecm,angle=0,file=xiloetas.eps}}
\vspace{4mm}
\centerline{\psfig{width=10truecm,angle=0,file=xiloetao.eps}}
\vspace{2mm}
\caption{ $R_s\equiv\xi_s/L$ versus $\eta_s$ (above) and $R_o\equiv \xi_o/L$
  versus $\eta_o$ (below).  The full lines are the results of the random
  spin-wave theory, obtained by using the expressions (\protect\ref{G-RSWT}).
  The MC data are obtained for the CRPXY and GRPXY models with $\sigma=0.1521$.  } 
\label{xiloeta} 
\end{figure}

The above-reported results allow us to evaluate the exponents
$\eta_s$ and $\eta_o$ which are related to the large-distance 
behavior of the spin and overlap correlation functions: 
\begin{eqnarray}
&&G_s(r)\sim r^{-\eta_s},
\qquad \eta_s=
     \cases{ {T+\sigma\over 2\pi} & (GRPXY) \cr 
             {T+\sigma+\sigma^2/2\over 2\pi} & (CRPXY)} , 
\label{etasw}\\
&&G_o(r)\sim r^{-\eta_o},
\qquad \eta_o={T \over \pi}.
\label{etaswo}
\end{eqnarray}
We can thus write for both models, at this level of approximation, 
\begin{eqnarray}
G_s(x-y) = \exp[2\pi \eta_s G(x-y)] \qquad 
G_o(x-y) = \exp[2\pi \eta_o G(x-y)].
\label{G-RSWT}
\end{eqnarray}
Note that the disorder dependence is completely included in the exponents 
$\eta_s$ and $\eta_o$.
Using these expressions we can campute the universal functions
$R_s(\eta_s)$ and $R_o(\eta_o)$, which express the ratios $R_s\equiv
\xi_s/L$ and $R_o\equiv \xi_o/L$  in terms of the corresponding exponents
$\eta_s$ and $\eta_o$. It is clear that $R_s(\eta_s)$ and $R_o(\eta_o)$ are 
identical [$R_s(x)=R_o(x)$] and disorder independent, hence they coincide with those
relevant for the pure XY model. These functions are shown in Fig.~\ref{xiloeta}.
Below we show numerically that these predictions are satisfied by our numerical data. 
This provides a clear evidence for the spin-wave nature of the
QLRO in the RPXY models.  A similar strategy was applied in \cite{HPV-05}
to clarify the nature of the low-temperature phase of fully frustrated
XY models.

\section{Monte Carlo results}
\label{mcres}

In this section we numerically investigate the nature of the QLRO phase that occurs
for sufficiently small disorder. In particular,
we want to provide a stringent check of the random spin-wave scenario.
The quantities which we compure are defined in \ref{notations}.

\subsection{Numerical details}
\label{mcsim}

We performed MC simulations of the GRPXY and CRPXY models,
considering square lattices of linear size $L$ with periodic boundary 
conditions. We set in both cases
$\sigma=0.1521$, which is well below the maximum value 
$\sigma_M$ ($\sigma_M\gtrsim 0.30$ in the CPRXY model \cite{APV-09}
and $\sigma_M\simeq \pi/8$ in the GRPXY model) and considered several values of $T$ 
below the critical temperature $T_c(\sigma)$, which marks the end
of the QLRO phase.  MC simulations in the
high-temperature phase~\cite{APV-09} indicate $T_c=0.771(2)$
for the GRPXY and $T_c=0.763(1)$ for the CRPXY. 

In the simulations we used a mixture of standard Metropolis 
and overrelaxed microcanonical updates: A single MC step
consisted of five microcanonical sweeps followed by 
one Metropolis sweep.  In the MC simulations of the CRPXY model we also used
the parallel-tempering method \cite{raex,par-temp}. It allowed us to perform 
efficient simulations in the region $T\lesssim \sigma$.  
In the parallel-tempering simulations
we considered $N_T$ systems at the same value of $\sigma$ and
at $N_T$ different inverse temperatures $\beta_{\rm min} \equiv \beta_1$,
\ldots, $\beta_{\rm max}$. The largest value $\beta_{\rm max}$ 
corresponded to the minimum
temperature value we were interested in.  The value $\beta_{\rm min}$ was
chosen in the paramagnetic phase and was such that thermalization at 
$\beta=\beta_{\rm min}$ was sufficiently fast.
The intermediate values $\beta_i$ were chosen such that the acceptance
probability of the temperature exchange was at least 5\%. 
Morever, we always  included the value $\beta = 1/\sigma$, which lies along
the N line. This provided a check of the numerical programs, since the 
MC results could be compared with the known exact results~\cite{Nishimori-02}.
The overlap correlation functions and corresponding $\chi_o$ and $\xi_o$
were obtained by simulating two independent replicas for each disorder sample. 

In the case of the GRPXY model we performed standard MC simulations at
$T=2/3,\,1/2,\,2/5$, for lattice sizes $10\lesssim L\lesssim 40$. 
Typically, we considered 50000 disorder realizations and performed 
$O(10^5)$ MC steps for each of them.
In the case of the CRPXY model the parallel-tempering method allowed us to
investigate the temperature range $T<T_c$ down to $T\approx 0.139$, which is below
the N line, i.e. satisfies $T<\sigma=0.1521$.  We performed simulations for lattice
sizes $10\lesssim L\lesssim 30$. Typically, we considered 25000 disorder realizations.
For $\beta\equiv 1/T=7.2$ we also
performed standard MC runs up to $L=85$.

\subsection{QLRO in RPXY models}
\label{lowtph}

\begin{figure}[tb]
\centerline{\psfig{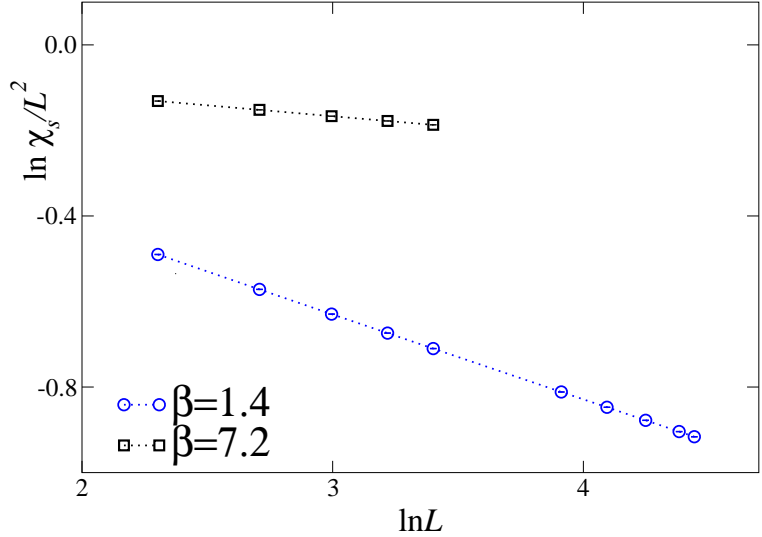}}
\vspace{2mm}
\caption{
MC estimates of $\ln \chi_s/L^2$  versus $\ln L$ for the CRPXY 
model at $\sigma=0.1521$ and two values of 
$\beta\equiv 1/T$, corresponding to 
$T\approx 0.139$ and $T\approx 0.714$.
}
\label{chivl}
\end{figure}

\begin{figure}[tb]
\centerline{\psfig{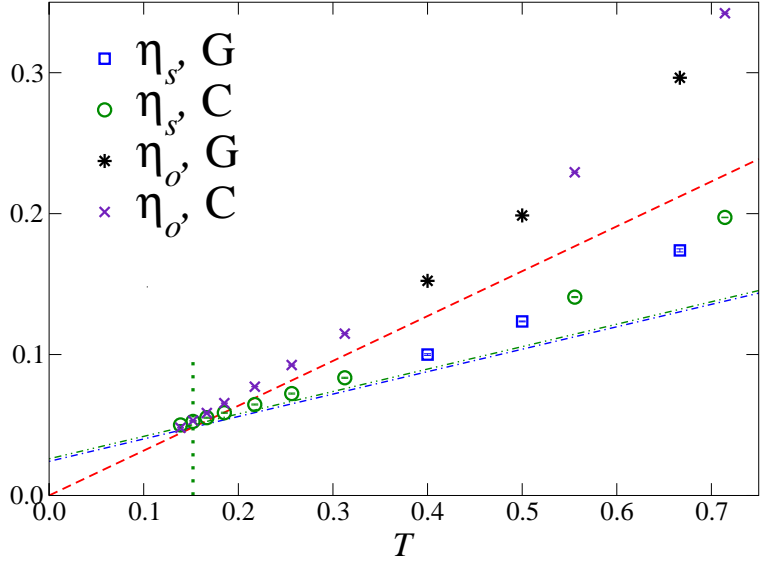}}
\caption{
The MC estimates of $\eta_s$ and $\eta_o$ vs $T$
for the GRPXY and CRPXY models at $\sigma=0.1521$.
The lines shows the spin-wave approximations
(\ref{etasw}), and (\ref{etaswo}). The two lines that 
give $\eta_s$ for the GRPXY and CRPXY models 
cannot be easily distinguished on the scale of the plot.
The dotted vertical line corresponds to 
$T=\sigma=0.1521$; in the CRPXY model this point belongs to the 
N line. 
}
\label{etavst}
\end{figure}

\begin{figure}[tb]
\centerline{\psfig{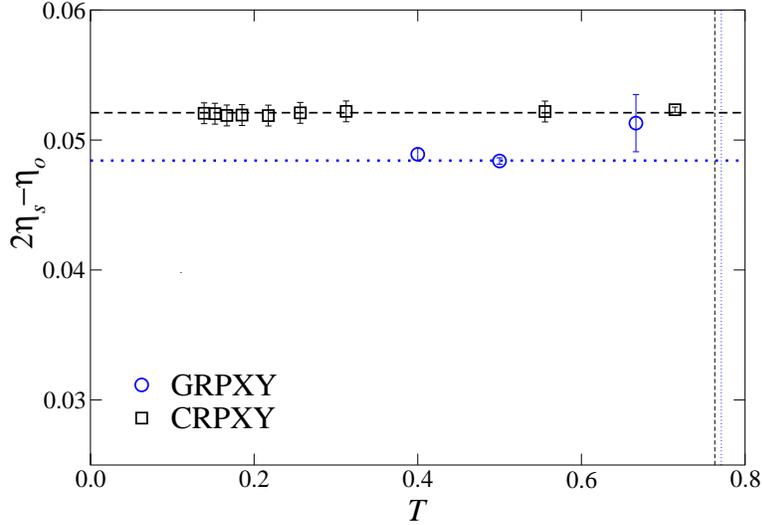}}
\vspace{2mm}
\caption{ We plot the difference $2 \eta_s - \eta_o$ vs $T$
  at $\sigma=0.1521$, and compare it with the low-order spin-wave
  approximations for the GRPXY and CRPXY models, cf.~(\ref{testdiffG})
  and (\ref{testdiffC}), respectively dotted and dashed lines. The vertical
  dotted and dashed lines indicate the critical temperatures of the two
  models, i.e. $T_c=0.771(2)$ and $T_c=0.763(1)$ for the GRPXY and CRPXY,
  respectively.}
\label{diffeta}
\end{figure}

We estimate the exponents $\eta_s$ and $\eta_o$ 
by studying the finite-size behavior of the
susceptibilities $\chi_s$ and $\chi_o$ defined in \ref{notations}.
Indeed, for $L\to \infty$ they behave as
\begin{equation}
\chi_{s,o}  \sim L^{2-\eta_{s,o}}~.
\label{chiso}
\end{equation}
Estimates of $\chi_s$ are shown in Fig.~\ref{chivl}.
On a logarithmic scale, the data fall on a straight line, 
indicating that the asymptotic behavior (\ref{chiso}) already
holds for the values of $L$ we consider.
In Fig.~\ref{etavst} we show the estimates of
$\eta_s$ and $\eta_o$ for the GRPXY
and CRPXY model. For $T\lesssim 0.2$ they agree with the spin-wave
approximations (\ref{etasw}) and (\ref{etaswo}). Morever, they appear to be 
mostly independent of the model, in agreement with the 
random spin-wave predictions (note that $\sigma^2/2 = 0.0116$ for 
$\sigma = 0.1521$).
For a more quantitative check, in
Fig.~\ref{diffeta} we plot the difference $2 \eta_s - \eta_o$ vs $T$,
and compare it with the low-order approximations
\begin{eqnarray}
&2 \eta_s - \eta_o = {\sigma\over \pi} \quad & \quad {\rm (GRPXY)},
\label{testdiffG}\\
&2 \eta_s - \eta_o = {\sigma + \sigma^2/2\over \pi} 
\quad & \quad {\rm (CRPXY)}.
\label{testdiffC}
\end{eqnarray}
The agreement is very good. Morever, 
the above-reported relations appear to hold up to temperatures
close to the KT transition $T_c$ (for $\sigma=0.1521$ we have 
$T_c=0.771(2)$ for the GRPXY model and
$T_c=0.763(1)$ for the CRPXY model), suggesting that they
may also hold at the KT transition.  Since MC simulations in the
high-temperature phase~\cite{APV-09} show a clear evidence that $\eta_s=1/4$,
this may suggest that at the KT transition
\begin{eqnarray}
\eta_o\approx {1\over 2} - {\sigma\over \pi} \quad & \quad {\rm (GRPXY)},\\
\eta_o\approx {1\over 2} - {\sigma+\sigma^2/2\over \pi}
\quad & \quad {\rm (CRPXY)},
\end{eqnarray}
The most important check of the spin-wave nature of the QLRO is provided by
the MC data shown in Figs.~\ref{xiloeta}, where we plot $R_s$ vs $\eta_s$ and
$R_o$ vs $\eta_o$: they agree with high accuracy with the curves $R_s(\eta_s)$
and $R_o(\eta_o)$ computed in the spin-wave limit.  We believe that these
results provide a conclusive evidence that the QLRO phase is determined by
random spin-wave theory.

\begin{figure}[tb]
\centerline{\psfig{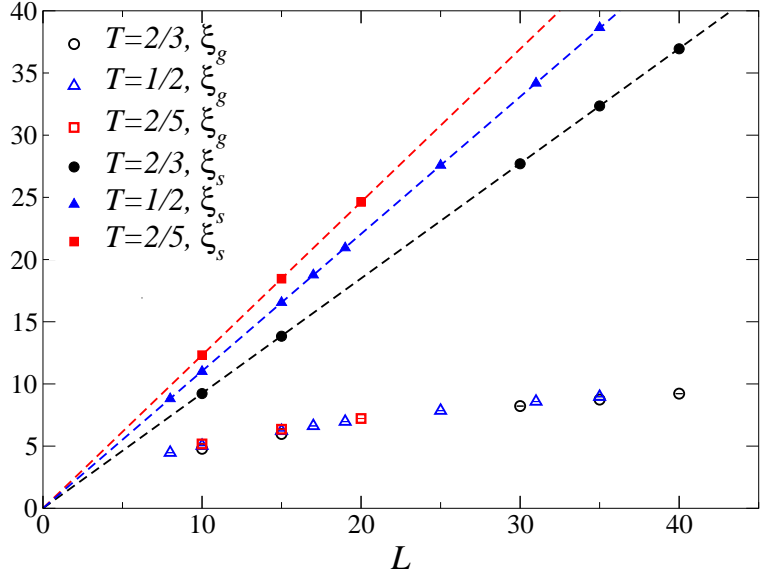}}
\vspace{2mm}
\caption{ The correlation lengths $\xi_s$, defined from the standard spin-spin
  correlation function, and $\xi_g$ defined from a gauge invariant spin-spin
  correlation function, versus $L$.  Results for the GRPXY model with
  $\sigma=0.1521$.  }
\label{ximg}
\end{figure}

Finally, we report some results for the {\em gauge-invariant}
correlation function (\ref{gaugecorr}), see \ref{notations}.  
In the spin-wave limit one
finds~\cite{Nishimori-02}
\begin{equation}
[\langle e^{i(\phi_x-A_{x,x'}-...A_{y',y}-\phi_y)} \rangle] 
= {\rm exp}[(T-\sigma)G(x-y)-|x-y|\sigma/2],
\label{sw2gg} 
\end{equation}
which predicts that gauge-invariant spin-spin correlation functions are not
critical. For instance,
in the large-$L$ limit the correlation length $\xi_g^{\rm (gap)}$ 
defined from the large-distance exponential decay of the 
gauge-invariant correlation function is finite and given by 
$\xi_g^{\rm (gap)}=2/\sigma$, independently of $T$. 
These predictions are confirmed by our MC simulations. 
We compute the second-moment correlation function
$\xi_g$ defined in~(\ref{smc}). The results are reported in 
Fig.~\ref{ximg}. It is evident that 
$\xi_g$ is finite in the large-$L$ limit, satisfies 
$\xi_g\lesssim 2/\sigma \simeq 13 $, and is independent of $T$.
Again, this result shows that the critical behavior is correctly described by
the spin-wave theory.

\section{Conclusions} \label{conclusions}

In this paper we have studied the low-temperature low-disorder phase 
in RPXY models. We have considered two different disorder distributions
and for each of them we have computed numerically 
the exponents $\eta_s$ and $\eta_o$, and the correlation lengths $\xi_s$ and 
$\xi_o$. These results have been compared with the predictions 
of the random spin-wave theory. Our main results are the following.
\begin{itemize}
\item[1)]
The ratios $\xi_s/L$ and $\xi_o/L$, 
when expressed in terms of the corresponding exponents $\eta_s$ and $\eta_o$,
are in perfect agreement with the analytic predictions. This indicates 
that the expressions (\ref{G-RSWT}) hold quite precisely in the 
whole low-temperature QLRO phase . 
\item[2)] 
The expressions (\ref{etasw}) and (\ref{etaswo}) hold only for very low
values of $T$. However, the difference $2\eta_s - \eta_o$ is 
apparently well described by spin-wave theory up to the 
critical transition which marks the end of the paramagnetic phase.
\item[3)] 
In agreement with the random spin-wave theory, the gauge-invariant 
spin correlation function (\ref{gaugecorr}) is not critical.
\end{itemize}
Finally, note that our calculations refer to probability distributions 
for which $[A] = 0$. Very little changes if we consider a 
nonzero average; for instance, one might use the distribution
\begin{equation}
P(A) \propto \exp\left[-{(A-a)^2/2\sigma}\right]~.
\end{equation}
In this case we have $[A]=a$. 
The new model can be mapped into the original one by performing 
the gauge transformation
\begin{equation}
\psi_{(x_1,x_2)}' = e^{-i a (x_1+x_2)}\psi_{x_1,x_2} \qquad \qquad 
A_{ij}' = A_{ij} - a .
\label{anotzero-gf}
\end{equation}
Hence this model has the same phase diagram as the original one.
The trasformation (\ref{anotzero-gf}) 
leaves the overlap correlation functions unchanged, 
since they are gauge invariant. The behavior of the 
spin-spin magnetic correlation functions is more subtle. 
If $b = (a,a)$, in Fourier space we have 
\begin{equation}
\widetilde{G}_s(q;a) = \widetilde{G}_s(q+b;a=0),
\end{equation}
where $\widetilde{G}_s(q;a)$ is the Fourier transform of the spin magnetic
correlation function in the theory with a nonvanishing average $a$. 
In the standard theory ($a=0$) the critical modes are those with $q=0$, 
while for $q\not=0$ the behavior is not critical. This implies that the 
critical modes in the theory with $a\not=0$ are those associated with
a nonvanishing momentum $q=-b$. Hence, in this theory 
the magnetic susceptibility, which corresponds to $q=0$, is not critical.

As a final comment, we note that the distribution functions of the phase
shifts are not gauge invariant: Phase shifts that differ only by a gauge
transformation have different probabilities.  Another interesting issue is
whether the results for the QLRO phase reported here also apply to
gauge-invariant distributions.

\appendix

\section{Notations}
\label{notations}

In terms of complex site variables $\psi_i\equiv
e^{i\theta_i}$, the RPXY Hamiltonian takes the form
\begin{equation}
{\cal H} = -\sum_{\langle ij \rangle } {\rm Re} \,\psi_i^* U_{ij} \psi_j
\label{h2}
\end{equation}
where $U_{ij}\equiv e^{i A_{ij}}$.

We consider several two-point correlations functions: the magnetic spin-spin
correlation function\footnote{The last equality in 
(\ref{magcorr}) can be proved by using the symmetry 
$\psi_x\rightarrow {\psi}_x^*$ and
$U_{xy}\rightarrow {U}_{xy}^*$.}
\begin{equation}
G_s(x-y) \equiv {\rm Re}\, [ \langle {\psi}_x^* \,\psi_y \rangle ]=
[ \langle {\psi}_x^* \,\psi_y \rangle ],
\label{magcorr}
\end{equation}
and the overlap correlation function
\begin{equation}
G_o(x-y) \equiv [ |\langle {\psi}_x^* \,\psi_y \rangle|^2 ],
\label{overcorr}
\end{equation}
which can be written as $G_o(x-y) = [ \langle
\bar{q}_x \,q_y \rangle ]$, where $q_x =
{\psi}_x^{(1)*} \psi_x^{(2)}$ and the upperscripts refer to two independent
replicas with the same disorder.
The angular and square  brackets indicate
the thermal average and the quenched average over disorder,
respectively.  We also consider a gauge-invariant spin-spin 
correlation function
\begin{equation}
G_g(x-y) \equiv [ {\rm Re} \langle {\psi}_x^* 
\,U[\Gamma_{x;y}] \,\psi_y \rangle ],
\label{gaugecorr}
\end{equation}
where $\Gamma_{x;y}$ is a path that connects sites $x$ and $y$ and
$U[\Gamma_{x;y}]$ is a product of phases associated with the links that belong
to $\Gamma_{x;y}$.  The paths connecting the points $x$ and $y$ are chosen
along the lattice axes, choosing the shortest path (see \cite{APV-08} for
details).

We define the corresponding susceptibilities: the magnetic
susceptibility $\chi_s\equiv \sum_x G_s(x)$, the overlap susceptibility
$\chi_o\equiv \sum_x G_o(x)$, and $\chi_g\equiv \sum_x G_g(x)$. We also define
the corresponding second-moment correlation lengths
\begin{equation}
\xi_{\#}^2 \equiv {\widetilde{G}_\#(0) - 
\widetilde{G}_\#(q_{\rm min}) \over 
          \hat{q}_{\rm min}^2 \widetilde{G}_\#(q_{\rm min}) },
\label{smc}
\end{equation}
where $q_{\rm min} \equiv (2\pi/L,0)$, $\hat{q} \equiv 2 \sin q/2$, and
$\widetilde{G}_\#(q)$ is the Fourier transform of $G_\#(x)$, and $\#$
indicates $s,o,g$.

\section{Random spin-wave computation of $\eta_s$} \label{App:etasCPRXY}

In this Appendix we wish to derive~(\ref{GsCPRXY}).  We follow closely
\cite{RSN-83}.  We first consider the spin-spin correlation function
$G_s(r,A_\alpha) = \langle \exp\left[i (\phi(0) - \phi(r))\right] \rangle$ for
fixed values of the random phases $A_\alpha$.  As in \cite{RSN-83} we rewrite
it as
\begin{equation}
G_s(r) = {
   \langle \exp\left[i (\phi(0) - \phi(r)) - \beta \int d^2 r A\cdot \nabla \phi\right] 
    \rangle_0 \over 
   \langle \exp\left[ - \beta \int d^2 r A\cdot \nabla \phi\right] \rangle_0
   },
\end{equation}
where $\langle \cdot \rangle_0$ indicates the average with Hamiltonian 
\begin{equation}
{\cal H} =  {\beta\over2} \int d^2 r (\nabla \phi)^2~.
\label{SW-H}
\end{equation}
Repeating the steps discussed in \cite{RSN-83} we end up with
\begin{equation}
G_s(r,A) = \exp \left( T G(r) + {1\over2} \int d^2 s \sum_\alpha A_\alpha(s) M_\alpha(s,r)
    \right),
\end{equation}
where $G(r)$ is the Gaussian propagator (\ref{Gaussian-prop}) and
\begin{equation}
M_\alpha(s,r) = 2 
    \int {d^2q\over (2\pi)^2} e^{-iq\cdot s} (e^{iq\cdot r} - 1) {q_\alpha \over q^2}~.
\end{equation}
Note that $M_\alpha(s,r)$ is imaginary, $M_\alpha(s,r)^* = - M_\alpha(s,r)$, so that 
\begin{equation}
|G_s(r,A)|^2 = e^{2 T G(r) }~.
\end{equation}
Thus, irrespective of the phase distribution, the overlap correlation function
does not depend on $\sigma$. To compute the spin correlation function
we must average $G_s(r,A)$ over the distribution of the phases $A_\alpha$. We 
consider the general distribution
\begin{equation}
P(A) \propto \exp \left(- {Q(A^2)\over 2\sigma}\right),
\end{equation}
which satisfies 
$Q(z) = z$ for $z\to 0$ and is such that, for $\sigma\to 0$ the distribution is peaked 
around $A = 0$. Thus, to compute the expansion of $G_s(r)$ for small $\sigma$,
we can expand $Q(A^2)$ in powers of $A^2$. We assume 
\begin{equation}
Q(z) = z + \alpha z^2 + O(z^3),
\end{equation}
where $\alpha$ is a distribution-dependent coefficient. For the 
distribution (\ref{paxy2}) we have $\alpha = -1/12$. To compute the correction of 
order $\sigma^2$ to $\eta_s$ we rewrite 
($[\cdot]_A$ indicates the average over $A$) 
\begin{eqnarray}
S &\equiv &
 \left[ \exp \left( {1\over2} \int d^2 s \sum_\alpha A_\alpha(s) M_\alpha(s,r)
    \right) \right]_A  \nonumber \\
  &\propto&  \int [DA] 
   \exp \left( - {1\over 2\sigma} 
       \int d^2 s [A^2 + \alpha (A^2)^2] + 
    {1\over2} \int d^2 s \sum_\alpha A_\alpha M_\alpha~\right).
\end{eqnarray}
We introduce a new field $B_\alpha$ defined by 
\begin{eqnarray}
   A_\alpha = \sqrt{\sigma} B_\alpha + {\sigma\over2} M_\alpha
\end{eqnarray}
and perform the integral over $B$. Disregarding terms of order $\sigma^3$ 
we obtain 
\begin{equation}
S = \exp \left[{\sigma\over8} \left(1 - 6 \alpha \sigma\right)
    \int d^2 s\, M(s,r)^2 \right].
\end{equation}
Since 
\begin{equation}
   \int d^2 s\, M(s,r)^2 = 8 G(r) ,
\end{equation}
we obtain finally
\begin{equation}
G_s(r) = e^{(T + \sigma - 6 \alpha \sigma^2) G(r)}.
\end{equation}
If we set $\alpha = -1/12$, we obtain result (\ref{GsCPRXY}).

\end{document}